\documentclass[cameraready]{Interspeech}

\title{\textsc{Latent-Mark}: An Audio Watermark Robust to Neural Codec Compression}

\author[affiliation={1,2},equalcontribution]{Yen-Shan}{Chen}
\author[affiliation={1,3, 4},equalcontribution]{Shih-Yu}{Lai}
\author[affiliation=1]{Ying-Jung}{Tsou}
\author[affiliation=1]{Yi-Cheng}{Lin}
\author[affiliation=1]{\\Bing-Yu}{Chen}
\author[affiliation=1]{Yun-Nung}{Chen}
\author[affiliation={1,5}]{Hung-yi}{Lee}
\author[affiliation=1,correspondingauthor]{Shang-Tse}{Chen}

\address{
  $^1$National Taiwan University, Taiwan \  
  $^2$CyCraft AI Lab, Taiwan \\
  $^3$RIKEN Center for Computational Science (RIKEN-CCS), Japan \\
  $^4$MoonShine Animation Studio, Taiwan \\
  $^5$NTU Artificial Intelligence Center of Research Excellence (NTU AI-CoRE)
}

\email{ \{r14922018, r13922a22, r14922076\}@csie.ntu.edu.tw, f12942075@ntu.edu.tw,
robin@ntu.edu.tw, 
y.v.chen@ieee.org, 
hungyilee@ntu.edu.tw, 
stchen@csie.ntu.edu.tw }

\keywords{Audio Watermarking,
Neural Codec Compression,
Latent-Space Shift,
Manifold Alignment,
Cross-Codec Transferability}

\usepackage{comment}

\usepackage{microtype}
\usepackage{graphicx}
\usepackage{subcaption}
\usepackage{booktabs} 
\usepackage{bbm}
\usepackage{multirow}
\usepackage{enumitem}
\usepackage{xcolor}
\newcommand{\acc}[3]{#1 {\scriptsize \color{gray}(#2/#3)}}

\usepackage{hyperref}
\usepackage{paralist}
\usepackage[numbers]{natbib}
\usepackage{amsmath,amssymb,amsfonts}
\usepackage{mathtools}

\usepackage{graphicx}   
\usepackage{makecell}   

\begin{document}

\maketitle

\begin{abstract}
    While existing audio watermarking techniques have achieved strong robustness against traditional digital signal processing (DSP) attacks, they remain vulnerable to neural compression. This occurs because modern neural audio codecs act as noise filters and discard the imperceptible waveform variations used in prior watermarking methods. To address this limitation, we propose Latent-Mark, the first zero-bit audio watermarking framework designed to survive \update{neural codec} compression. Our key insight is that robustness to the encode-decode process requires embedding the watermark within the codec's invariant latent space. We achieve this by optimizing the audio waveform to induce a detectable directional shift in its encoded latent representation, while constraining perturbations to align with the natural audio manifold to ensure imperceptibility. To prevent overfitting to a single codec's quantization rules, we introduce Cross-Codec Optimization, jointly optimizing the waveform across multiple surrogate codecs to target shared latent invariants. Extensive evaluations demonstrate robust zero-shot transferability to unseen neural codecs, achieving competitive resilience against traditional DSP attacks while preserving perceptual imperceptibility.
    We hope our work will inspire future research into universal watermarking frameworks capable of maintaining integrity across increasingly complex and diverse generative distortions.
    \footnote{All source code and implementation details are available at: \url{https://github.com/yenshan0530/Latent-Mark}}
\end{abstract}
\section{Introduction}
\begin{figure}[t!] 
    \centering
    \begin{subfigure}[b]{0.9\linewidth}
        \centering
        \includegraphics[width=\linewidth]{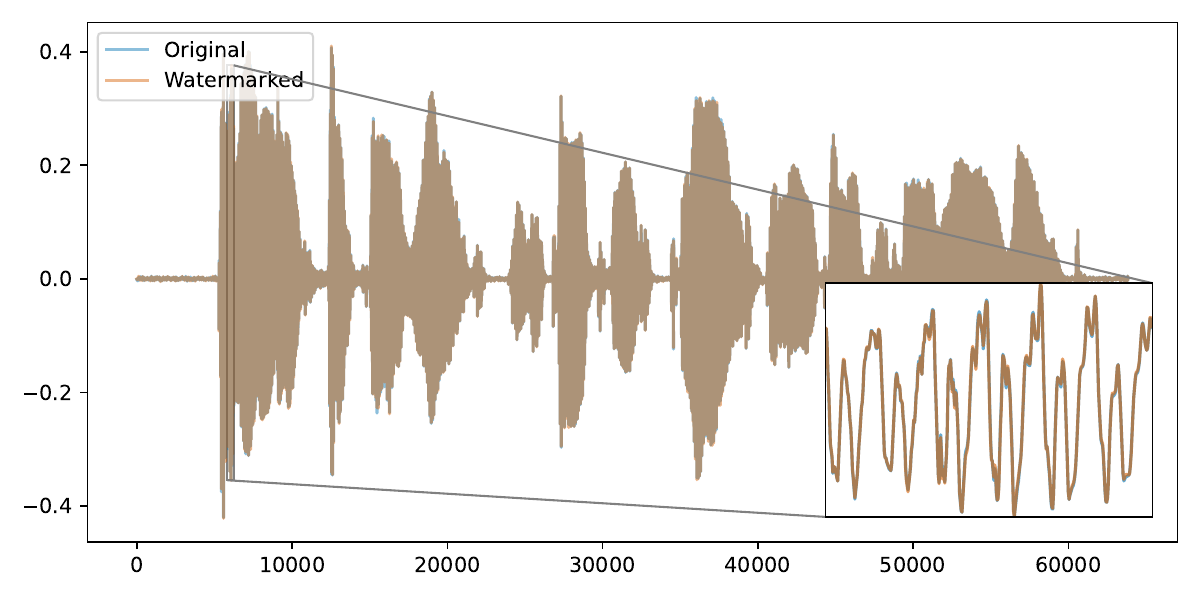}
        \caption{Original vs. Watermarked}
        \label{fig:orig-wm}
    \end{subfigure}
    

    \begin{subfigure}[b]{0.9\linewidth}
        \centering
        \includegraphics[width=\linewidth]{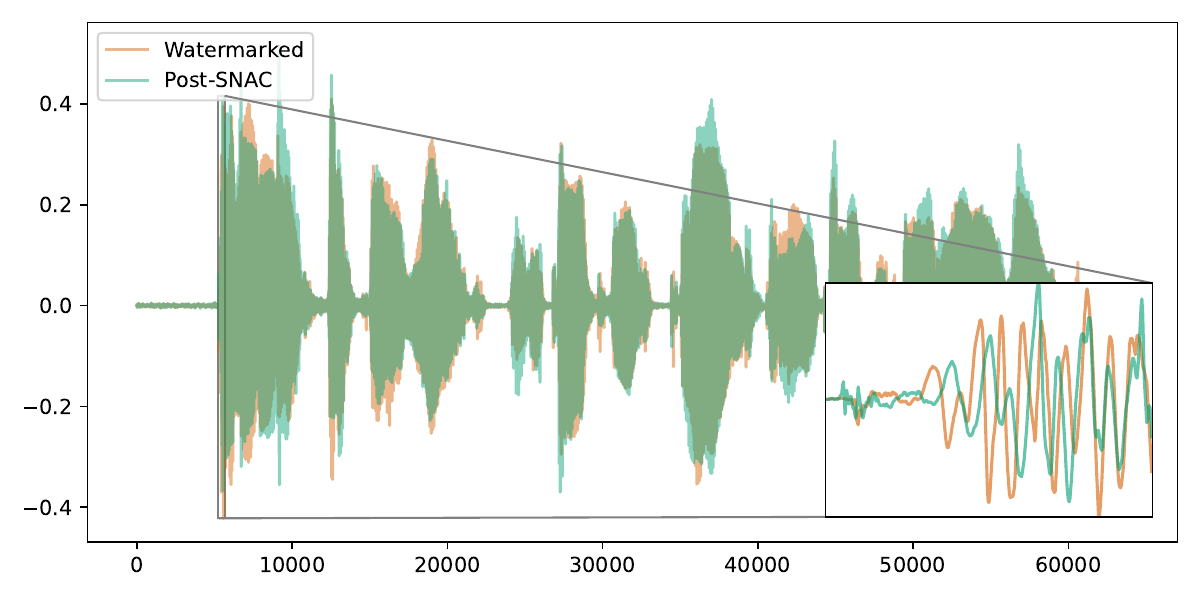}
        \caption{Watermarked vs. Post-SNAC}
        \label{fig:wm-snac}
    \end{subfigure}

    \caption{
    Waveform comparisons at different processing stages using AudioSeal. 
    (a) The original and watermarked waveforms largely overlap, showing minimal differences. 
    (b) Substantial amplitude distortion and phase shifts occur after the SNAC encoding--decoding process.
    }
    \label{fig:waveform-preliminary}
\end{figure}

Audio watermarking has emerged as a critical tool for intellectual property protection. Recent state-of-the-art methods, such as AudioSeal~\cite{audioseal_2024}, WavMark~\cite{wavmark_2024}, and Timbre~\cite{timbre_2023}, demonstrate strong resilience against a broad spectrum of conventional digital signal processing (DSP) distortions~\cite{wen2025sok}, including compression, filtering, resampling, and sample suppression. In such settings, watermark robustness has largely been framed as surviving waveform- or spectrogram-level perturbations while remaining imperceptible to human listeners.

However, the rapid adoption of neural network-based codec compression introduces a qualitatively different threat surface that breaks this traditional notion of robustness. Modern neural audio codecs, such as EnCodec~\cite{encodec} and SNAC~\cite{snac}, reconstruct audio by mapping waveforms into discrete latent tokens and decoding them back under a strict bitrate constraint. This encode--quantize--decode pass, which we refer to as \textbf{Neural Codec Compression}, is not a localized DSP distortion but a highly non-linear projection through a learned latent bottleneck. As a result, existing watermarks that are highly robust to DSP transformations can fail catastrophically after a single codec pass. We investigated this impact through a preliminary analysis using AudioSeal~\cite{audioseal_2024}, the SNAC~\cite{snac} codec, and the LibriSpeech~\cite{librispeech} dataset. As illustrated in Figure~\ref{fig:orig-wm}, while a watermarked waveform initially exhibits no visible deviation from the original signal, a single neural encode--decode pass (Figure~\ref{fig:wm-snac}) introduces extensive phase shifts and amplitude distortions that completely misalign with the source watermark.

We attribute this behavior to the inherent design of neural codecs, which act as 
 a \update{projector}
 mapping input signals onto the latent space of valid audio representations. Under this projection, traditional watermarks---typically embedded as imperceptible
noise---are treated as off-manifold residuals and discarded during reconstruction. Despite the output remaining perceptually transparent, these structural modifications effectively strip away the fine-grained signals of watermarks required for detection. This elevates neural codec compression from a routine compression task to a potent watermark removal attack, representing a pressing real-world vulnerability as codec-based pipelines become the de facto standard for generative modeling and audio distribution.

To withstand this regime, we propose Latent-Mark, the first \emph{zero-bit} audio watermarking framework (i.e., it encodes only \emph{presence} rather than a payload) explicitly designed to withstand neural codec compression.
Our core insight is that only features embedded in a codec's invariant latent space can survive its encode--quantize--decode process. Accordingly, we formulate watermark embedding as a latent-targeted optimization problem: we apply gradient-based updates directly to the input waveform to induce a detectable \emph{directional shift} in the codec latent space, while constraining waveform perturbations so that the watermark remains imperceptible. By embedding the mark into the latent manifold itself, it becomes a feature the codec is designed to preserve, rather than a waveform-level artifact neural codecs are trained to discard.

A practical watermark must satisfy more than surviving neural codec distortions.
\textbf{First}, it must remain imperceptible. We ensure this by constraining the latent shift to align with the codec's learned representation space—specifically along directions defined by codebook centroids.
This alignment leverages the decoder to naturally regularize the watermark, preserving acoustic fidelity.
\textbf{Second}, it must generalize beyond a single codec. Building on the above white-box embedding strategy, we further introduce cross-codec optimization across diverse surrogate codecs. By enforcing the watermark objective
simultaneously under multiple quantization rules, the framework captures shared \update{latent patterns}
across different codecs and avoids overfitting to a single architecture. This ensures zero-shot transferability to unseen black-box codecs. 
\textbf{Finally}, the watermark should remain robust to traditional DSP distortions: in addition to neural codec compression, we evaluate Latent-Mark and find that it maintains state-of-the-art resilience against a wide range of analytic attacks, including adding Gaussian noise, scaling amplitude, filtering, and resampling.



Our primary contributions are as follows:
\begin{compactitem}
    \item We identify \textbf{neural codec compression} as a fundamentally different attack regime for audio watermarking, and argue that neural codecs act as \update{manifold projectors}
    that can erase the imperceptible
    noise patterns used by prior watermarking methods.
    \item We propose Latent-Mark, the first zero-bit audio watermarking framework explicitly designed to withstand neural codec compression by inducing a detectable \textbf{latent directional shift} via gradient-based waveform optimization. We show that \textbf{latent space-aligned} shifts enhance acoustic \textbf{imperceptibility} (Sections~\ref{sec:direction_analysis}, ~\ref{sec:imperceptibility}).
    \item Building on the white-box formulation, we introduce cross-codec optimization across multiple codecs to achieve strong zero-shot \textbf{transferability} to unseen black-box codecs. Ultimately, Latent-Mark provides a balance of perceptual fidelity and survivability to 
    \update{neural codec compression}, while maintaining highly competitive \textbf{robustness against prior DSP attacks} (Section~\ref{sec:dsp_attack}).
\end{compactitem}

\section{Related Work}

\subsection{Mechanisms of Neural Watermarking}
Across audio, image, text, and multi-modal domains, watermarking methods are categorized by the space in which they embed signals \citep{liu2024audiomarkbench, ozer2025rawbench, zhou2024wmcodec, zhang2024zodiac, transfer_attack_image_wm_2024, diaa2024adaptive_attacks, chen2024demark, rastogi2025stamp, durability_multimodal_benchmark_2024, vlamark_2025}. A recurring theme is that robustness depends on how watermark signals interact with model-internal representations and survive distortions. To formalize this, we categorize several core concepts commonly employed in the design of robust watermarking algorithms:

\vspace{5pt}
\noindent
\textbf{Post-Hoc Audio Signal and Additive Perturbations.}
While traditional watermarking treats signals as additive perturbations across modalities \citep{transfer_attack_image_wm_2024}, audio presents a uniquely stringent setting due to the ubiquity of lossy compression and model-mediated transformations \citep{vall-e2023}. In contrast to image or text modalities, audio watermarks face the unique challenge of surviving the complex digital signal processing (DSP) inherent to audio production workflows, followed by subsequent degradation across distribution channels. Specifically, they face repeated neural codec re-synthesis and quantization (e.g., SNAC \citep{snac}, APCodec \citep{APCodec}, and FunCodec \citep{FunCodec}, etc.), which RAW-Bench \citep{ozer2025rawbench} identifies as the most formidable challenge to bit-string integrity. Furthermore, Özer \emph{et al.} \citep{ozer2025rawbench} emphasize that practical audio watermarks must withstand an extensive battery of studio-standard manipulations—including \textit{dynamic range compression, limiting, equalization, and reverberation}—alongside environmental degradation like background noise. Audio watermarking methods like AudioSeal \citep{audioseal_2024}, SilentCipher \citep{singh2024silentcipher}, Timbre \citep{timbre_2023}, and WavMark \citep{wavmark_2024} address these vulnerabilities by incorporating psychoacoustic masking and noise-to-mask ratio losses \citep{noise_to_mask_ratio_2024} to hide information within inaudible frequency bands.


\vspace{5pt}
\noindent
\textbf{Latent and Manifold-Aware Embedding.}
Recognizing that deep generative models manipulate semantic concepts rather than raw signals, recent work has shifted toward embedding marks directly into continuous latent spaces or discrete codec tokens \citep{zhang2024zodiac, slic_2024, zhou2024wmcodec}. Building on classical manifold learning \citep{tenenbaum2000isomap, roweis2000lle} and vector-quantized representations \citep{oord2017vqvae}, these methods inject structured perturbations into deep features to find a subspace separable from natural content variations. In this paradigm, a watermark’s survival depends on aligning with model-internal structural invariances—such as pitch or speaker identity—rather than superficial signal details \citep{sadok2025interpretability, tokui2025latentgranular}. To ensure robustness against the quantization and re-synthesis inherent in model-mediated transformations, current strategies include training-time codec augmentation \citep{juvela2025codecaug} or integrating watermarking objectives directly into neural codecs via end-to-end joint training \citep{zhou2024wmcodec}. This motivates evaluating watermark schemes explicitly under latent representations, such as those used in multi-modal or diffusion-based frameworks \citep{durability_multimodal_benchmark_2024, vlamark_2025}.

\vspace{5pt}
\noindent
\textbf{Audio Latent-Space Approaches and Their Limitations.} Most recently, highly concurrent works have attempted to tackle the neural codec bottleneck by operating near or within latent spaces, though they rely on fundamentally different paradigms. For instance, Roman \emph{et al.} \citep{LatentGen} explore watermarking the \emph{training data} of audio generative models such that the trained model inherently emits watermarked codec tokens. However, this assumes white-box control over the model training pipeline and fails to address the \emph{post-hoc} marking of arbitrary audio assets in the wild. Alternatively, Liu \emph{et al.} \citep{XAttnMark} propose an end-to-end framework utilizing cross-attention mechanisms for robust embedding. While effective against known distortions, it relies on training a static, feed-forward neural encoder against a predefined attack set. This risks overfitting to the specific quantization rules seen during training, fundamentally limiting its \emph{zero-shot transferability} to unseen, proprietary neural codecs. Furthermore, without explicit geometric constraints, modifying latent representations can easily push the signal off the natural audio manifold, introducing perceptible artifacts upon decoding. 
In contrast, our Latent-Mark overcomes these limitations. Rather than training a static encoder or altering generative models, we employ test-time \textit{Cross-Codec Optimization} across surrogate codecs to induce a measurable directional shift in shared \update{latent}
structures. This ensures zero-shot survival against black-box neural codec compression while enforcing natural manifold constraints to guarantee imperceptibility.

\subsection{Watermark Robustness and the Threat of Distortions}
The definition of a ``robust'' watermark has evolved significantly, shifting from algorithmic signal degradations to complete \update{audio}
reconstruction.

\vspace{5pt}
\noindent
\textbf{Traditional DSP, Channel Distortions, and Robustness.}
Robustness is increasingly defined by rigorous evaluation protocols that transition from algorithmic signal-level corruptions to complex \update{audio}
reconstructions. Historically, benchmarks like AudioMarkBench \citep{liu2024audiomarkbench} and RAW-Bench \citep{ozer2025rawbench} systematized a wide array of removal and forgery attacks, including MP3 compression, bandpass filtering, added noise, reverberation, and modern neural codec chains. To counter these, researchers have developed various defensive mechanisms, such as noise-to-mask ratio (NMR) losses \citep{noise_to_mask_ratio_2024}, training-time codec augmentation \citep{juvela2025codecaug}, and end-to-end deepfake verification pipelines \citep{waveverify_2025}. Beyond generic perturbations, contemporary systems must withstand adaptive and model-aware threats, including overwriting, ownership conflicts, and spoofing \citep{yours_or_mine_overwriting_2025}. These security challenges have prompted the integration of defensive cryptographic hashing and filter-based mechanisms to prevent forgery within neural watermarking frameworks \citep{hashed_watermark_filter_2025}.

\vspace{5pt}
\noindent
\textbf{The Paradigm Shift to Neural Codec Compression.} The landscape of distortions fundamentally changed with the introduction of high-fidelity neural audio codecs (e.g., SoundStream \citep{zeghidour2021soundstream}, EnCodec \citep{encodec}, DAC \citep{kumar2023dac}). Unlike MP3, which removes psychoacoustically masked frequencies, neural codecs completely deconstruct the waveform into discrete 
tokens---typically utilizing convolutional architectures and Residual Vector Quantization (RVQ) to capture coarse-to-fine structure---and re-synthesize it from scratch. Because neural codec compression acts as an extreme, structure-aware information bottleneck \citep{wen2025sok, oreilly2025deepaudiowatermarksshallow, ozer2026selfvoiceconversionattack}, it explicitly discards the ``off-manifold'' residual noise that traditional additive methods rely upon. Compounding this threat, these discrete codec tokens now serve as the foundational vocabulary for modern generative sequence models \citep{borsos2022audiolm, agostinelli2023musiclm, copet2023musicgen}. Large audio language models treat quantized indices as linguistic sequences for zero- or few-shot speech synthesis \citep{vall-e2023, vall-e-2-2025, le2023voicebox, rubenstein2023audiopalm, peng2024voicecraft}. Thus, signal-space modifications must survive this encode--quantize--decode pipeline. Recent benchmarks like RAW-Bench \citep{ozer2025rawbench} confirm neural codecs are the primary threat to audio watermarks, necessitating a shift toward latent-aware strategies.

\begin{figure*}[t]
     \centering
      \includegraphics[width=0.9\linewidth]{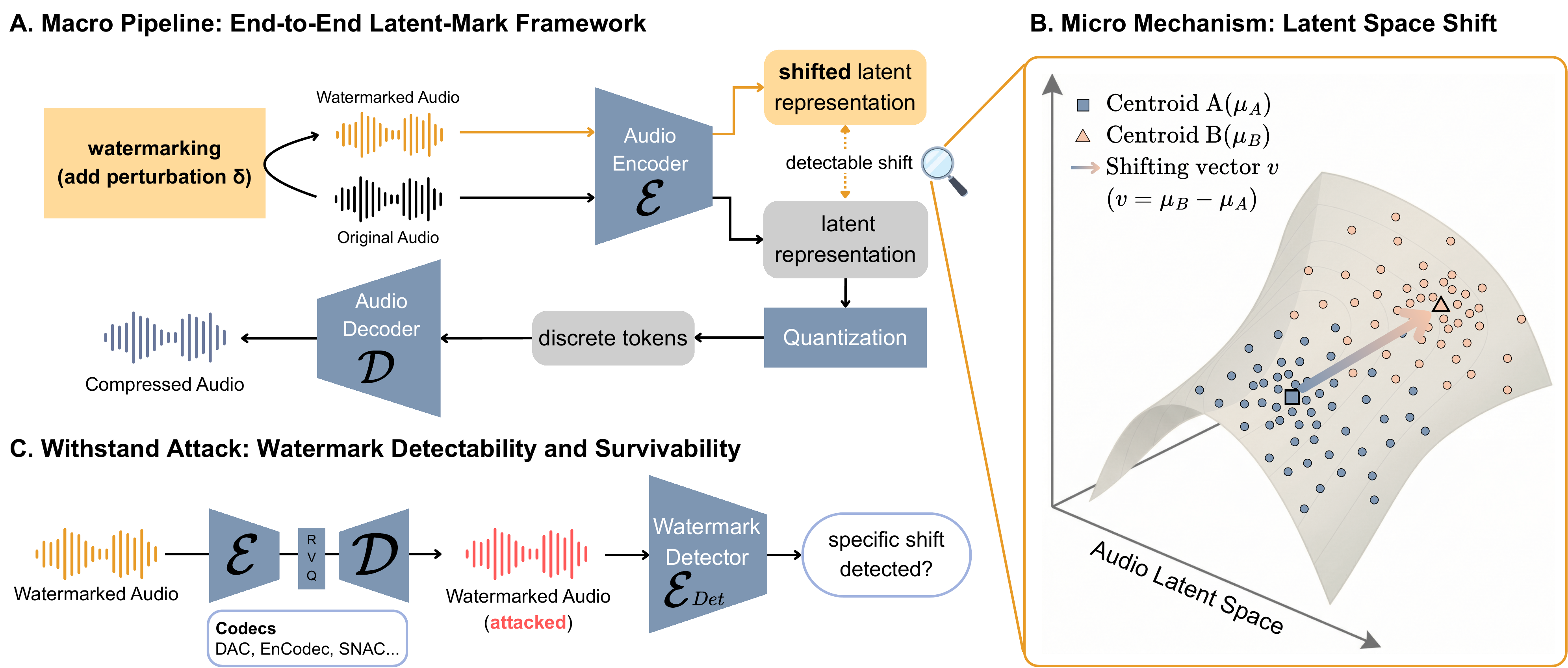}
     \caption{
     \textbf{Overview of Latent-Mark.} (A) \textbf{Macro} end-to-end pipeline. The lower blue and gray components illustrate the standard neural codec compression pipeline; The orange components highlight our proposed additions: adding an optimized perturbation $\delta$ to the original audio to intentionally induce a constrained shift in its latent representation prior to quantization and decoding. (B) \textbf{Micro} mechanism for the robust latent space shift, driven by a predefined shifting vector $v = \mu_{B} - \mu_{A}$ between cluster centroids. (C) Watermark detectability and survivability against \textbf{attacks}. The watermarked audio undergoes destructive neural codec pipelines (encoding, RVQ, and decoding), after which a detector ($\mathcal{E}_{Det}$) verifies if the latent shift persists.
     } 
     \label{fig:conceptual}
\end{figure*}

\section{Methodology}

\subsection{Preliminaries}

\textbf{The Gap.} The vulnerability of traditional audio watermarking to neural codec compression stems from a fundamental \textbf{representation mismatch}. Conventional methods typically embed watermarks as
psychoacoustic masking patterns at the waveform level. However, modern neural codecs process audio through a lossy reconstruction operator:
\begin{equation}
\mathcal{R}(s) = \mathcal{D}(\mathcal{Q}(\mathcal{E}(s)))
\end{equation}
where $\mathcal{E}$ is the encoder mapping the time-domain audio $s$ to a continuous feature space; $\mathcal{Q}$ is the vector quantizer that discretizes these features by mapping them to the nearest entries in a codebook $\mathcal{W}$; and $\mathcal{D}$ is the decoder that reconstructs the signal back into the original audio space. Because traditional watermarks are designed to be imperceptible, they often manifest as subtle signal variations that are effectively treated as quantization noise. During codec compression, these ``off-manifold'' details are discarded.
Consequently, the core challenge stems from \textbf{detectability}: the basic ability to extract the watermark from an unattacked signal, and \textbf{survivability}: the capacity of the watermark to endure the lossy quantization of codec compression.


\vspace{5pt}
\noindent
\textbf{The intuition.} Assuming a white-box setting where we have parameter access to the codec's encoder $\mathcal{E}$, we propose Latent-Mark: a framework (Figure ~\ref{fig:conceptual}) that identifies and leverages latent properties invariant to the quantization process. Rather than injecting waveform-level noise, we induce a directional shift in the continuous latent representation $z$ toward a secret manifold axis $v_c$ before it enters the quantizer $\mathcal{Q}$. This modification is designed to survive codec compression by ``steering'' the latent tokens into a distribution with a persistent directional bias. While the perceived audio remains unchanged, this shift remains detectable even after the signal is decoded and subsequently re-encoded. We formulate this as a \textbf{zero-bit watermarking challenge}, where the goal is to induce a statistically significant shift in the latent sequence along a designated secret axis, ensuring the perturbation survives the \update{neural compression}
bottleneck.


\subsection{Problem Formulation}
\label{sec:formulation}

Let $s \in \mathbb{R}^T$ be the input audio waveform, where $T$ denotes the total number of temporal samples. A neural codec $c$ maps $s$ to a latent representation:
\begin{equation}
z_c = \mathcal{Q}(\mathcal{E}(s)) \in \mathbb{R}^{d \times L}
\end{equation}
where $d$ is the codebook dimension and $L$ is the sequence length of the latent frames. Our framework consists of an Encoder $\mathcal{E}_{wm}$ and a Detector $\mathcal{E}_{Det}$. We generate a watermarked waveform $s_{wm} = s + \delta$ by solving a gradient-based optimization over the perturbation $\delta$. The Detector determines the presence of the mark by computing a scalar score $y$ based on the latent projection of a suspect signal $s'$ onto a secret manifold vector $v_c \in \mathbb{R}^d$.

To induce the desired shift while maintaining perceptual transparency, we formulate the embedding as a constrained optimization problem. We solve for the perturbation $\delta$ that maximizes the alignment with the manifold vector $v_c$ subject to a small waveform distortion:
\begin{equation}
\min_{\delta} \mathcal{L}_{\text{wm}}(s + \delta, v_c) \quad \text{s.t.} \quad ||\delta||_{\infty} \le \epsilon 
\end{equation}
where $\epsilon$ is a dynamic threshold determined by the target Signal-to-Distortion Ratio (SDR). This prevents the watermark from introducing audible artifacts into the original audio $s$.

The Latent-Mark framework consists of an optimization-based Encoder and a statistical Detector. For any given codec $c$, the presence of a watermark is defined by the alignment of the signal's latent representation $z_c$ with a secret manifold axis $v_c \in \mathbb{R}^d$. We measure this alignment via the projection score $\bar{p}_c(s)$, defined as the temporal mean of the latent projections over the sequence length $L$:
\begin{equation}
\bar{p}_c(s) = \frac{1}{L} \sum_{t=1}^{L} \langle z_{c,t}, v_c \rangle 
\end{equation}

\noindent
\textbf{Watermark Optimization.} Embedding is formulated as a constrained optimization problem. Given an input audio $s$, we solve for a waveform perturbation $\delta$ that steers the latent representation into the target direction. We minimize a hinge loss objective:
\begin{equation}
\update{\min_{\delta} \mathcal{L}_{\text{wm}}(s + \delta, v_c) = \min_{\delta} \text{ReLU}(\gamma_c - \bar{p}_c(s+\delta))}
\end{equation}
where $\gamma_c$ represents the \textbf{target alignment score} (set to $1.5$ in our implementation). The intuition behind $\gamma_c$ is to create a ``safety margin'' that pushes the latent projection far enough into the target manifold so that the shift survives the rounding effects of the quantization bottleneck. To maintain imperceptibility, we enforce a dynamic threshold $\epsilon = \beta \cdot \text{RMS}(s) \cdot 10^{-\text{SDR}/20}$, where RMS denotes the root mean square function. We solve this via the Adam optimizer for 150 steps, applying a hard clip to $\delta$ at each iteration \update{for imperceptibility}.

\vspace{5pt}
\noindent
\textbf{Watermark Detection.} Detection is a statistical verification process performed in the latent domain. For a suspect signal $s'$, we first compute the raw projection score $\bar{p}_c(s')$, which corresponds to the temporal mean of the projected latent sequence. To ensure robustness across different codec architectures with varying latent scales, we compute a \textbf{Normalized Margin} $m_c$:
\begin{equation}
m_c(s') = \frac{\bar{p}_c(s') - \tau_c}{\sigma_c} 
\end{equation}
where $\tau_c$ is the detection threshold ($\mu_c + k\sigma_c$) and $\sigma_c$ is the standard deviation of projections derived from a null distribution of clean audio. A watermark is detected if $m_c > 0$.

\vspace{5pt}
\noindent

\noindent
\textbf{Choice of Shifting Axis.} While $v_c$ can be any arbitrary unit vector, \lily{for the main experiment,} we design its selection with the intention of improving survivability through the quantization bottleneck $\mathcal{Q}$\lily{, and term this the \textbf{Latent-Cluster} variant}. Aiming to guide the shift toward high-density regions of the latent space, we derive $v_c$ by partitioning the codebook weights $W \in \mathbb{R}^{K \times d}$ into two primary groupings via $K$-means clustering ($k=2$). Letting $\mu_A$ and $\mu_B$ be the resulting centroids, we define the axis as the unit-normalized vector between them:
\begin{equation}
v_c = \frac{\mu_B - \mu_A}{||\mu_B - \mu_A||_2} 
\end{equation}

By aligning $v_c$ with the codebook distribution, our intention is for the perturbation to act more like a structural feature rather than stochastic noise, hypothesizing that this alignment increases its likelihood of preservation during discretization. A comparative analysis of alternative directions is in Section \ref{sec:direction_analysis}.

\subsection{Cross-Codec Optimization}
\label{sec:joint-opt}

A fundamental limitation of single-codec optimization is the lack of transferability; a watermark optimized for one specific codec's latent representation may be treated as stochastic noise and discarded by another. To achieve zero-shot robustness against unseen black-box models, we introduce \textbf{Cross-Codec Optimization}. Instead of targeting a single bottleneck, we identify 
a directional shift that a committee of heterogeneous surrogate codecs $\mathcal{C}$ collectively deems structural.

\vspace{5pt}
\noindent
\textbf{Framework Overview.} Our joint optimization pipeline consists of four integrated stages: (1) a multi-rate resampling loop to synchronize heterogeneous codec views; (2) a calibration phase to balance gradients across different latent scales; (3) a constrained optimization loop that induces a structurally invariant latent shift; and (4) an ensemble detection mechanism that aggregates evidence for robust verification.

\vspace{5pt}
\noindent
\textbf{Stage 1. Cross-Codec Resampling Pipeline.} Because the committee members operate at different sampling rates, we implement a synchronized resampling loop. During each optimization step, the perturbation $\delta$ is maintained in a high-resolution workspace at rate $f_{\text{work}}$. For each codec $c \in \mathcal{C}$, the perturbed signal $s + \delta$ is resampled to the codec's native rate $f_c$ before latent extraction. Waveforms are padded to satisfy a temporal constraint $T_\text{pad}$ to ensure stable optimization across all views. In our implementation, we set $f_\text{work} = 44.1$\,kHz and $T_\text{pad}$ to multiples of 4096 samples.

\vspace{5pt}
\noindent
\textbf{Stage 2. Gradient Balancing via Calibration.} Different architectures operate on vastly different latent scales, which can lead to ``gradient dominance'' where one codec's loss overwhelms others. To counter this, we use a \textbf{Baseline Calibration} method to calculate a target threshold $\tau_c = \mu_c + k\sigma_c$ and a normalization scale $\alpha_c$ derived from a null distribution of clean audio:
\begin{equation}
\alpha_c = \mathbb{E} \left[ \text{ReLU}(\tau_c - \bar{p}_c(s)) \right] 
\end{equation}
where $\alpha_c$ represents the average projection gap of clean audio. We set the calibration constant $k = 1.5$. By dividing each per-codec hinge loss by its respective $\alpha_c$, we ensure the optimization assigns equal importance to satisfying the manifolds of all committee members regardless of their latent variance.


\vspace{5pt}
\noindent
\textbf{Stage 3. Cross-Codec Optimization.} We solve for the optimal perturbation $\delta$ by minimizing the ensemble normalized hinge loss over $N_{steps}$ using the Adam optimizer:
\begin{equation}
\label{eq:joint-loss}
\min_{\delta} \frac{1}{|\mathcal{C}|} \sum_{c \in \mathcal{C}} \frac{\text{ReLU}(\tau_c - \bar{p}_c(s+\delta))}{\alpha_c} \quad \text{s.t.} \quad ||\delta||_{\infty} \le \epsilon \end{equation}
The budget $\epsilon$ is dynamically adjusted as $\epsilon = \text{clip}(\beta \cdot \text{RMS}(s) \cdot 10^{-\text{SDR}/20}, \epsilon_{min}, \epsilon_{max})$. We set $N_{steps} = 150$, $\beta = 2.5$, $\epsilon_{min} = 10^{-4}$, and $\epsilon_{max} = 0.1$. This objective effectively induces a directional bias into the shared \update{latent}
features of the committee, increasing the likelihood that the mark will survive codec compression by an unseen attacker codec $a \notin \mathcal{C}$.


\vspace{5pt}
\noindent
\textbf{Stage 4. Ensemble Detection and $\Delta$-Score.} Detection is performed by aggregating evidence across the committee. For a suspect signal $s'$, we compute the \textbf{Normalized Margin} $m_c(s') = (\bar{p}_c(s') - \tau_c) / \sigma_c$ for each view. Let $\{ m_{(1)}, m_{(2)}, \dots, m_{(|\mathcal{C}|)} \}$ denote the margins sorted in ascending order. To ensure robustness against outlier distortions, the final detection score is defined as:
\begin{equation}
\label{eq:detect-joint}
\text{score}(s') = m_{\left( \left\lceil \frac{|\mathcal{C}|}{2} \right\rceil \right)}(s')
\end{equation}


Notably, taking the mean of normalized margins is susceptible to extreme outliers caused by scale variations across different codec geometries, while the median acts as a robust statistic. It effectively functions as a majority voting mechanism, ensuring that even if an attack completely obliterates the watermark in a minority of views (resulting in unbounded negative margins), the global detection remains stable as long as the structural bias persists in the remaining surrogate spaces.

To evaluate transferability under attack, we utilize a \textbf{$\Delta$-Score} metric, which measures the shift in detection score relative to a clean baseline: $\Delta = \text{score}(R_a(s_{wm})) - \text{score}(R_a(s))$, where $R_a$ is the codec compression operator of an unseen attack codec. A positive $\Delta$ indicates that the watermark's directional bias has successfully transferred through the black-box bottleneck.

\section{Experiments}
The primary objective of our study is to verify whether Latent-Mark retains its embedded signal through the extreme bottleneck of neural codec compression, and investigate if cross-codec joint optimization enables better transferability to unseen codecs.

\begin{table*}[t]
\centering
\caption{Benchmark results across datasets. Detectability (Det.) is presented as Accuracy (TPR/FPR), while Survivability (Sur.) reports the detection rate after neural codec compression with SNAC (24kHz). \zoe{For simplicity, the original version of our \textbf{Latent-Mark} is shown as \textbf{Latent-Cluster}, and \textbf{Latent-Mark-Joint} is abbreviated to \textbf{Latent-Joint}. }\textbf{Left:} Comparison of Latent-Mark and its joint-optimization variant against prior baselines (Section~\ref{sec:main_results}). \textbf{Right:} Evaluation of our method using alternative secret key directions (Section~\ref{sec:direction_analysis}). Best and second-best values are highlighted in bold and underlined.}
\label{tab:watermark_results}
\resizebox{\textwidth}{!}{%
\begin{tabular}{ll lll|ll ll}
\toprule
& & \multicolumn{5}{c}{\textit{Main Experiment}} & \multicolumn{2}{c}{\textit{Directional Variants}} \\
\cmidrule(lr){3-7} \cmidrule(lr){8-9}
Dataset & Metric & WavMark & SilentCipher & AudioSeal & Latent-Cluster & Latent-Joint & Latent-PCA & Latent-Random \\
\midrule
\multirow{2}{*}{AIR} 
& Det. & \acc{~~95.0}{99.2}{9.2} & \acc{~~93.3}{95.0}{9.2} & \acc{~~\underline{96.7}}{97.5}{4.2} & \acc{~~95.8}{95.8}{4.2} & \acc{\textbf{100.0}}{100.0}{0.0} & \acc{~~95.8}{99.2}{6.7} & \acc{\textbf{100.0}}{100.0}{0.0} \\
& Sur. & ~~~~0.0 & ~~~~0.0 & ~~~~0.0 & ~~\underline{61.7} & ~~53.3 & ~~60.8 & \textbf{~~66.7} \\
\midrule
\multirow{2}{*}{Clotho} 
& Det. & \acc{~~93.3}{95.0}{9.2} & \acc{~~\textbf{96.7}}{100.0}{5.8} & \acc{~~\underline{94.2}}{94.2}{5.8} & ~~\acc{92.5}{91.7}{7.5} & \acc{~~\textbf{96.7}}{95.8}{3.3} & \acc{~~93.3}{92.5}{7.5} & \acc{~~93.3}{95.0}{9.2} \\
& Sur. & ~~~~0.0 & ~~~~0.0 & ~~~~0.0 & ~~\underline{58.3} & ~~\underline{58.3} & ~~\textbf{61.7} & ~~\textbf{61.7} \\
\midrule
\multirow{2}{*}{DAPS} 
& Det. & \acc{\textbf{100.0}}{100.0}{0.0} & \acc{~~93.3}{97.5}{11.7} & \acc{\textbf{100.0}}{100.0}{0.0} & \acc{~~\underline{99.2}}{100.0}{1.7} & \acc{~~83.3}{85.0}{20.0} & \acc{~~95.0}{99.2}{9.2} & \acc{~~95.8}{96.7}{5.0} \\
& Sur. & ~~~~0.0 & ~~~~0.0 & ~~~~8.3 & \textbf{~~93.3} & ~~76.7 & ~~\underline{88.3} & 65.0 \\
\midrule
\multirow{2}{*}{LibriSpeech} 
& Det. & \acc{~~\underline{96.7}}{96.7}{3.3} & \acc{~~\underline{96.7}}{97.5}{4.2} & ~~\acc{91.7}{91.7}{7.5} & \acc{\textbf{100.0}}{100.0}{0.0} & ~~\acc{93.3}{95.0}{10.0} & ~~\acc{95.8}{95.0}{3.3} & ~~\acc{95.0}{95.0}{5.0} \\
& Sur. & ~~~~4.2 & ~~~~0.0 & ~~~~5.0 & ~~\textbf{80.8} & ~~74.2 & ~~65.8 & ~~\underline{76.7} \\
\midrule
\multirow{2}{*}{jaCappella} 
& Det. & \acc{~~\underline{96.7}}{96.7}{3.3} & \acc{\textbf{100.0}}{100.0}{0.0} & \acc{\textbf{100.0}}{100.0}{0.0} & \acc{\textbf{100.0}}{100.0}{0.0} & ~~\acc{95.0}{98.3}{8.3} & ~~\acc{95.8}{95.8}{4.2} & \acc{\textbf{100.0}}{100.0}{0.0} \\
& Sur. & ~~~~0.0 & ~~~~0.0 & ~~~~0.0 & ~~\underline{75.8} & ~~\underline{75.8} & ~~\textbf{77.5} & ~~70.8 \\
\midrule
\multirow{2}{*}{PCD} 
& Det. & ~~\acc{90.8}{90.8}{9.2} & \acc{~~\underline{96.7}}{94.2}{0.0} & ~~\acc{90.8}{91.7}{10.0} & ~~\acc{90.8}{90.8}{9.2} & ~~\acc{88.3}{91.7}{16.7} & ~~\acc{95.8}{94.2}{1.7} & \acc{\textbf{100.0}}{100.0}{0.0} \\
& Sur. & ~~~~0.0 & ~~~~0.0 & ~~~~1.7 & ~~\textbf{81.7} & ~~\underline{78.3} & ~~71.7 & ~~75.8\\
\midrule
\multirow{2}{*}{MAESTRO} 
& Det. & \acc{~~\underline{96.7}}{99.2}{5.0} & \acc{~~\underline{96.7}}{95.0}{0.8} & ~~\acc{95.0}{99.2}{9.2} & \acc{\textbf{100.0}}{100.0}{0.0} & ~~\acc{89.2}{95.0}{16.7} & ~~\acc{95.0}{98.3}{8.3} & ~~\acc{95.8}{95.0}{3.3} \\
& Sur. & ~~~~0.0 & ~~~~0.0 & ~~~~0.0 & ~~\underline{80.8} & ~~71.7 & ~~65.0 & ~~\textbf{83.3} \\
\midrule
\multirow{2}{*}{GuitarSet} 
& Det. & \acc{\textbf{100.0}}{100.0}{0.0} & \acc{\textbf{100.0}}{100.0}{0.0} & \acc{\textbf{100.0}}{100.0}{0.0} & ~~\acc{96.7}{95.0}{0.8} & ~~\acc{85.0}{90.0}{20.0} & ~~\acc{95.0}{95.0}{5.0} & \acc{~~\underline{97.5}}{100.0}{4.2} \\
& Sur. & ~~~~0.0 & ~~~~0.0 & ~~~~0.0 & ~~\textbf{86.7} & ~~68.3 & ~~\underline{85.0} & ~~61.7 \\
\bottomrule
\end{tabular}%
}
\end{table*}

\subsection{Experimental Setup}
\textbf{Datasets.} We evaluate our method across nine diverse datasets spanning three primary acoustic domains: \textit{ambient and environmental sound} (AIR \citep{air_jeub09a}, Clotho \citep{clotho_9052990}), \textit{human speech} (LibriSpeech \citep{librispeech}, DAPS \citep{daps_mysore_2014_4660670}), and \textit{music and vocals} (PCD \citep{PCD_TISMIR}, jaCappella \citep{jaCapella_ICASSP}, MAESTRO \citep{maestro_hawthorne2018enabling}, GuitarSet \citep{GuitarSet}, Freischuetz \citep{Freischuetz}). To ensure statistical consistency and balanced evaluation, we uniformly sample 120 instances from each dataset, randomly subsampling those that exceed this threshold.



\vspace{5pt}
\noindent
\textbf{Baselines.} We compare Latent-Mark against three state-of-the-art watermarking models: \textbf{WavMark} \citep{wavmark_2024}, \textbf{SilentCipher} \citep{singh2024silentcipher}, and \textbf{AudioSeal} \citep{audioseal_2024}. For the single-codec variant of Latent-Mark, we utilize SNAC as the primary surrogate model. For the cross-codec variant (\textbf{Latent-Mark-Joint}), we optimize across an ensemble of models with diverse architectures and sampling frequencies (SNAC 32\,kHz, DAC 16\,kHz, and DAC 44\,kHz), testing cross-family codecs transferability on APCodec \citep{APCodec} and FunCodec \citep{FunCodec}, as formulated in Section~\ref{sec:joint-opt}. An extended analysis of this surrogate model selection is provided in Section~\ref{sec:exp-transfer}.

\vspace{5pt}
\noindent

\vspace{5pt}
\noindent
\textbf{Evaluation Metrics.} We evaluate watermark performance using two primary metrics. \textbf{Detectability (Det.)} evaluates the detector's fundamental accuracy on clean, unperturbed audio. We assess watermark performance using a balanced test set comprising 60 watermarked and 60 unwatermarked (original audio) samples per dataset. Alongside overall classification accuracy, we explicitly report the True Positive Rate (TPR) and the False Positive Rate (FPR).
\textbf{Survivability (Sur.)} quantifies the robustness of the embedded signal against severe compression bottlenecks. It is defined strictly as the successful detection rate on the watermarked samples after they have been processed through the neural codec compression attack.

\vspace{5pt}
\noindent
\textbf{Attacker Models.} In these primary evaluations, we utilize the SNAC~\cite{snac} architecture (with frequency 24kHz) as the surrogate model for Latent-Mark's detector. While our main experiments focus on survivability against SN codec compression, we further investigate the zero-shot transferability of our embedded marks in Section~\ref{sec:exp-transfer}, where we test attacks against
other unseen neural codecs in different codec types and sampling rates. \update{We use default values for all hyperparameters.}

\subsection{Results}
\label{sec:main_results}
Table~\ref{tab:watermark_results} (Left) summarizes the performance of our proposed Latent-Mark variants against state-of-the-art watermarking baselines across eight diverse datasets. From these evaluations, we draw three primary findings:

First, in terms of \textbf{detectability} on unperturbed audio, both the prior baselines and our Latent-Mark variants (with or without joint optimization) maintain highly competitive performance, consistently exceeding 0.95 accuracy across most datasets. This confirms that Latent-Mark, similar to prior baselines, reliably triggers the detector under standard conditions.

Second, a stark contrast emerges regarding \textbf{survivability} against the neural codec bottleneck. While prior watermarking methods experience catastrophic failure---dropping to near-zero detection rates across the board---Latent-Mark robustly preserves the embedded signal, achieving survivability scores consistently above 0.58 and peaking at 0.93 on the DAPS dataset (using Latent-Cluster). We note that \textit{Latent-Joint} exhibits slightly lower performance compared to its single-codec counterparts; this is expected, as performing white-box optimization against multiple codecs simultaneously necessitates a trade-off in specialized robustness. Nevertheless, its performance remains highly competitive, maintaining accuracy rates above 0.58 across all tested configurations.

Finally, we observe that the audio domain influences retention. Speech and vocal datasets (e.g., DAPS, LibriSpeech) generally exhibit more stable survivability, frequently exceeding 0.70. In contrast, certain environmental noise distributions like AIR show slightly lower retention ranges (between 0.50 and 0.70). Notably, among the baselines, only AudioSeal demonstrated any marginal resilience to neural codec compression (retaining a mere 0.08 on DAPS and 0.05 on LibriSpeech), further underscoring the critical necessity of our latent-space formulation.

\section{Ablation and Analysis}


\lily{Having demonstrated the primary robustness and detectability of Latent-Mark under neural codec compression, we now present a series of analyses to validate our design. Our evaluation is twofold. First, we conduct \textbf{architectural ablations} to justify our core hyperparameter choices, specifically investigating the \emph{selection of secret key directions} within the latent space (Section~\ref{sec:direction_analysis}) and the \emph{impact of surrogate model combinations} (Section~\ref{sec:exp-transfer}) Second, we evaluate Latent-Mark against established \textbf{baselines} to compare the \emph{acoustic imperceptibility} (Section~\ref{sec:imperceptibility}) of the watermarks and verify that optimizing for neural codec bottlenecks does not inadvertently compromise \emph{robustness against traditional digital signal processing attacks} (Section~\ref{sec:dsp_attack}).}

\subsection{Choice of Secret Key Direction}
\label{sec:direction_analysis}

\lily{We first discuss how the direction of the secret key (see Section~\ref{sec:formulation}, \textbf{Choice of Shifting Axis}) influences watermarking performance and survivability. To isolate the impact of this geometric choice, we compare three distinct variants for deriving the target axis: \textbf{Latent-Cluster} (our primary method, which utilizes the vector connecting the centroids of a $k=2$ clustering of the codebook), \textbf{Latent-PCA}, which derives the axis from the first principal component (the direction of maximum variance) of the centered codebook weights using Singular Value Decomposition, and \textbf{Latent-Random}, which simply samples a uniformly random, unit-normalized vector within the latent dimension space.}

\lily{Comparing the post-codec compression survivability of these approaches in Table~\ref{tab:watermark_results} (Right), Latent-Cluster yields the most robust performance, ranking first in four of the eight datasets. Latent-Random follows closely, securing first place in three datasets and second place in one. Conversely, Latent-PCA consistently performs the worst across the evaluations.}



\lily{\update{We hypothesize that t}his hierarchy stems from how each strategy interacts with the quantization bottleneck $\mathcal{Q}$. \update{For example,} \textbf{Latent-Cluster} \update{may} explicitly guide the watermark towards \update{denser}
regions of the codebook; by mimicking a structural transition between mass centers, the perturbation is effectively preserved by the nearest-neighbor quantizer as a valid \update{latent}
shift. Conversely, the poor performance of \textbf{Latent-PCA} \update{may} reveal that shifting along the axis of maximum variance is detrimental. Because the first principal component represents the most continuous variations in the feature space, the quantizer likely treats such shifts as standard signal variance rather than a distinct structural feature, causing the watermark to be aggressively washed out during discretization.}

\lily{Further comparisons regarding the acoustic imperceptibility of the watermarks generated by these distinct directional methods are detailed in Section~\ref{sec:imperceptibility}.}




\begin{table*}[t]
\caption{Transferability under codec compression attacks across all optimization settings and 3 acoustic domain datasets. Values are survivability pass rates (\%). The best and second-best values per column are highlighted in \textbf{bold} and \underline{underlined}.}
\label{tab:transfer_full}
\centering
\resizebox{\textwidth}{!}{%
\begin{tabular}{@{}l|ccc|ccc|ccc|ccc|ccc@{}}
\toprule
& \multicolumn{3}{c|}{\textit{Ambient \& Environmental}} & \multicolumn{6}{c|}{\textit{Human Speech}} & \multicolumn{6}{c}{\textit{Music \& Vocals}} \\
\cmidrule(lr){2-4}\cmidrule(lr){5-10}\cmidrule(lr){11-16}
\multirow{2}{*}{\textbf{Method}} &
\multicolumn{3}{c|}{\textbf{Clotho}} & 
\multicolumn{3}{c|}{\textbf{LibriSpeech}} & \multicolumn{3}{c|}{\textbf{DAPS}} &
\multicolumn{3}{c|}{\textbf{PCD}} & \multicolumn{3}{c}{\textbf{jaCappella}} \\
\cmidrule(lr){2-4}\cmidrule(lr){5-10}\cmidrule(lr){11-16}
&
\rotatebox{75}{\makecell{SNAC44}} & \rotatebox{75}{\makecell{EnCodec48}} & \rotatebox{75}{\makecell{DAC24}} &
\rotatebox{75}{\makecell{SNAC44}} & \rotatebox{75}{\makecell{EnCodec48}} & \rotatebox{75}{\makecell{DAC24}} &
\rotatebox{75}{\makecell{SNAC44}} & \rotatebox{75}{\makecell{EnCodec48}} & \rotatebox{75}{\makecell{DAC24}} &
\rotatebox{75}{\makecell{SNAC44}} & \rotatebox{75}{\makecell{EnCodec48}} & \rotatebox{75}{\makecell{DAC24}} &
\rotatebox{75}{\makecell{SNAC44}} & \rotatebox{75}{\makecell{EnCodec48}} & \rotatebox{75}{\makecell{DAC24}} \\
\midrule
Latent-Joint (Opt\_C1) & ~~92.50 & ~~\underline{97.50} & ~~\underline{93.33} & \textbf{100.00} & ~~85.00 & \textbf{100.00} & ~~90.83 & \textbf{100.00} & \textbf{100.00} & ~~95.00 & \textbf{100.00} & ~~90.00 & ~~92.00 & \textbf{100.00} & ~~68.00 \\
Latent-Joint (Opt\_C2) & \textbf{100.00} & ~~\textbf{98.33} & ~~79.17 & ~~\underline{80.00} & \textbf{100.00} & ~~79.17 & \textbf{100.00} & \textbf{100.00} & ~~77.50 & \textbf{100.00} & \textbf{100.00} & \textbf{100.00} & \textbf{100.00} & \textbf{100.00} & ~~\underline{98.00} \\
Latent-Joint (Opt\_F1) & ~~79.17 & ~~95.83 & ~~57.50 & \textbf{100.00} & ~~\underline{91.67} & \textbf{100.00} & ~~\underline{98.33} & \textbf{100.00} & ~~69.17 & ~~77.50 & \textbf{100.00} & ~~\underline{97.50} & ~~\underline{96.00} & \textbf{100.00} & ~~88.00 \\
\midrule
Latent-PCA (SNAC24) & ~~69.00 & ~~63.17 & ~~83.67 & ~~66.83 & ~~28.00 & ~~69.17 & ~~84.00 & ~~89.83 & ~~\underline{99.33} & \textbf{100.00} & \textbf{100.00} & \textbf{100.00} & ~~71.20 & ~~58.00 & ~~76.00 \\
Latent-Cluster (SNAC24) & \underline{94.67} & ~~95.33 & ~~\textbf{97.16} & \textbf{100.00} & \textbf{100.00} & ~~\underline{97.50} & \textbf{100.00} & \textbf{100.00} & \textbf{100.00} & ~~\underline{97.50} & \textbf{100.00} & \textbf{100.00} & \textbf{100.00} & \textbf{100.00} & \textbf{100.00} \\
Latent-Random (SNAC24) & ~~66.00 & ~~55.33 & ~~75.33 & ~~76.83 & ~~54.17 & ~~83.33 & ~~88.50 & ~~53.67 & ~~99.00 & ~~~~2.50 & ~~~~0.00 & ~~10.00 & ~~77.20 & ~~44.00 & ~~\underline{98.00} \\
\midrule
Latent-Joint (Opt\_D1) & ~~60.00 & ~~55.80 & ~~59.20 & ~~65.67 & ~~70.80 & ~~73.33 & ~~79.20 & ~~\underline{95.80} & ~~72.50 & ~~37.50 & ~~37.50 & ~~50.00 & ~~60.00 & ~~\underline{62.00} & ~~80.00 \\
Latent-Joint (Opt\_D2) & ~~70.80 & ~~63.30 & ~~61.70 & ~~47.50 & ~~69.17 & ~~44.17 & ~~90.80 & \textbf{100.00} & ~~82.50 & ~~55.00 & ~~\underline{52.50} & ~~52.50 & ~~40.00 & ~~44.00 & ~~56.00 \\
\bottomrule
\end{tabular}%
}
\end{table*}

\subsection{Surrogate Model Selection for Joint Optimization}
\label{sec:exp-transfer}

\lily{We next discuss how the model combination used for joint optimization (Section~\ref{sec:joint-opt}) affects performance and transferability.}

\lily{Throughout this section, codec shorthand (e.g., \texttt{SNAC32}) denotes the architecture (\texttt{SNAC}) operating at a specific sampling frequency (\texttt{32} kHz). We select five optimization combinations targeting two orthogonal generalization axes: \textit{cross-codec} architecture shifts and \textit{cross-sampling-rate} variations.}

\begin{compactitem}
    \item \lily{\textbf{Opt\_C1 \{SNAC32, DAC16, DAC44\}} and \textbf{Opt\_C2 \{SNAC32, EnCodec24, EnCodec32\}} focus on \textbf{intra-family} generalization. By optimizing against ``architecturally \underline{c}loser'' relatives sharing a hierarchical RVQ structure, we test if the watermark generalizes to unseen members of the same lineage across varying sampling rates.}

    \item \lily{\textbf{Opt\_F1 \{SNAC24, DAC24, EnCodec24\}} evaluates \textbf{cross-family} transferability under a controlled \underline{f}requency constraint by fixing all codecs at \texttt{24}\,kHz.}

    \item \lily{\textbf{Opt\_D1 \{SNAC24, DAC44, FunCodec\}} and \textbf{Opt\_D2 \{SNAC24, FunCodec, APCodec\}} explore extreme \textbf{cross-family} scenarios. These ``\underline{d}istant'' groups incorporate heterogeneous architectures (e.g., FunCodec, APCodec) to stress-test robustness against radical domain shifts in unrelated neural synthesis pipelines.}
\end{compactitem}

\lily{We compare these against single-optimization baselines (Latent-Cluster, PCA, and Random, optimized on \textbf{SNAC24}). To test transferability, we use \texttt{SNAC44}, \texttt{EnCodec48}, and \texttt{DAC24} for neural codec compression.}



\lily{Experimental results are summarized in Table~\ref{tab:transfer_full}. We define \textbf{transferability} as the success rate of watermark perturbations—initially optimized to survive a specific codec compression—when evaluated against unseen codec architectures. Specifically, we first identify watermark perturbations that successfully survive \texttt{SNAC24} codec compression (the common optimization target across all settings) and subsequently assess whether this robustness generalizes to disparate codec environments. We present results for two representative datasets from each audio category: environmental noise, human speech, and music. Several key observations can be drawn from these results:}

Architectural proximity serves as the primary determinant of transfer success, as evidenced by the significant performance gap between intra-family and cross-family transfers. Specifically, configurations optimized for architectures similar to the target (e.g., $C_1, C_2, F_1$) outperform distant-family transfers by approximately 20\%. While identifying shared \update{latent}
features across disparate architectures remains a challenge, our method maintains a baseline transferability between 50--70\%, indicating that the learned perturbations are not strictly overfitted to a single decoder's bias. Furthermore, intra-family generalization is significantly increased by including at least one representative codec from the target family during optimization. For instance, the inclusion of DAC variants in $C_1$ leads to high robustness for the unseen \texttt{DAC24}, while EnCodec-inclusive optimization in $C_2$ excels on \texttt{EnCodec48}, suggesting that structural commonalities, such as shared Residual Vector Quantization (RVQ) hierarchies, are more critical for transfer than matching bitrates or sampling frequencies.

\lily{The joint optimization configurations consistently demonstrate superior robustness compared to single-variant baselines, provided that the optimization set includes at least one representative from the target codec's architectural family. By comparing the joint frameworks ($C_1, C_2, F_1$) against the single-variant baselines (\textit{Latent-Cluster}, \textit{PCA}, and \textit{Random}), we observe that exposure to diverse neural codec compression processes prevents the watermark from occupying narrow, codec-specific latent regions, effectively covering a broader spectrum of unseen codecs than the individual
strategies.}




\lily{Finally, structural architecture remains a more significant bottleneck for watermark survival than sampling frequency, a fact highlighted by the $F_1$ configuration. Despite $F_1$ being optimized for 24kHz targets, it fails to provide disproportionate gains for the 24kHz \texttt{DAC24} variant compared to other families. This indicates that the specific inductive bias of the neural synthesis layers—rather than the frequency response of the signal—is the dominant factor that the watermark must navigate to remain imperceptible yet detectable.}

\begin{figure*}[t!]
     \centering
     
     \includegraphics[width=0.8\textwidth]{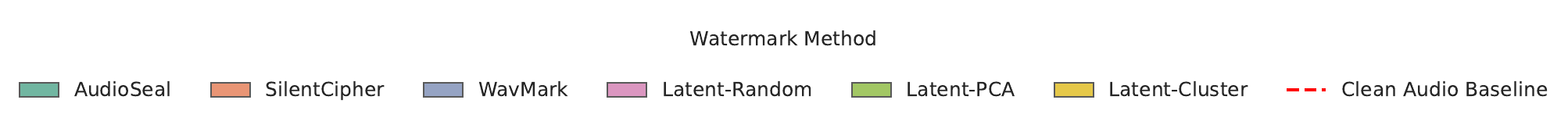} \\
    \vspace{-2mm}
     \begin{subfigure}[b]{0.6\textwidth} 
         \centering
         \includegraphics[height=4.5cm]{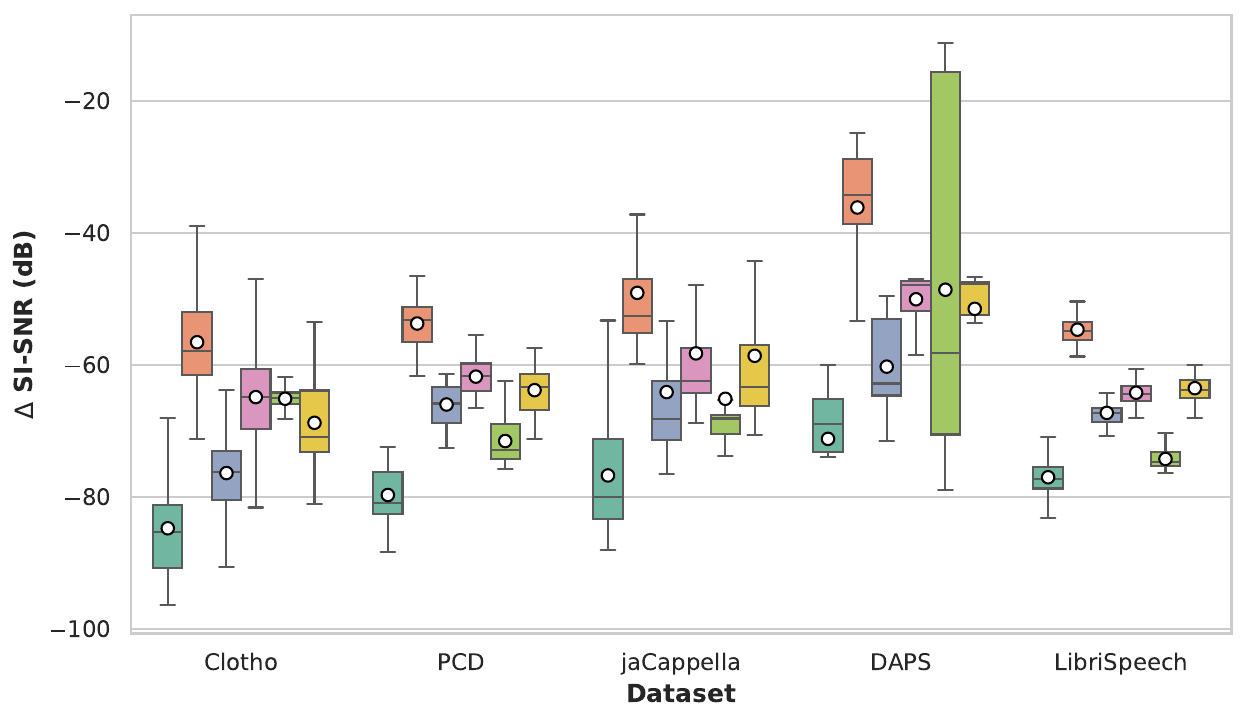}
         \caption{$\Delta$SI-SNR across datasets.}
         \label{fig:delta_sisnr}
     \end{subfigure}%
     \hfill
     \begin{subfigure}[b]{0.4\textwidth}
         \centering
         \includegraphics[height=4.5cm]{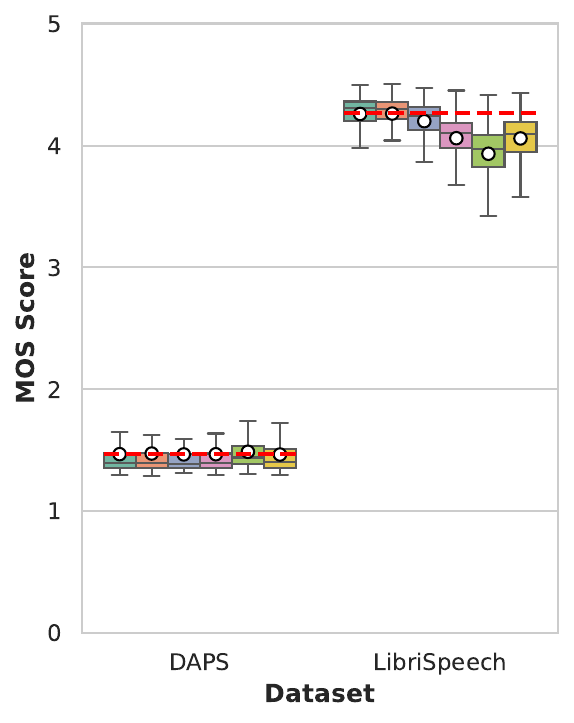}
         \caption{UTMOS on DAPS and LibriSpeech.}
         \label{fig:utmos}
     \end{subfigure}
    \vspace{-5mm}
     \caption{Comparison of objective waveform fidelity ($\Delta$SI-SNR) and perceptual quality (UTMOS) across watermarking methods.}
     \label{fig:quality_comparison}
\end{figure*}

\subsection{Audio Quality and Imperceptibility}
\label{sec:imperceptibility}

\lily{To evaluate the impact of watermark embedding on perceptual quality, we utilize two metrics: \textbf{$\Delta$SI-SNR} (Scale-Invariant Signal-to-Noise Ratio change)~\cite{luo2019sisnr} to measure mathematical waveform fidelity, and \textbf{UTMOS} (UTokyo-SaruLab MOS Prediction System)~\cite{saeki2022utmos}, a neural-based Mean Opinion Score (MOS) predictor for human-like perceptual assessment.}

\lily{As illustrated in Figure~\ref{fig:delta_sisnr}, the observed \textbf{change in waveform fidelity ($\Delta$SI-SNR)} follows the trend: \textit{SilentCipher} $>$ \textit{Latent-Cluster, Latent-Random} $>$ \textit{Latent-PCA}, \textit{AudioSeal}, \textit{WavMark}. Regarding our proposed variants, \textit{Cluster} and \textit{Random} demonstrate comparable stability, likely due to their uniform treatment of the latent space. In contrast, the PCA variant exhibits higher variance, reflecting greater sensitivity to the underlying data.}

\lily{As for \textbf{perceptual quality (UTMOS)}, despite the differences captured by $\Delta$SI-SNR, the UTMOS results in Figure~\ref{fig:utmos} indicate that the perceptual quality across all methods is nearly indistinguishable, suggesting that the artifacts introduced by \update{Latent-Mark}
are effectively masked and maintain a high level of imperceptibility to the human ear.}

\begin{table}[t!]
\caption{Survivability (pass rates, \%) under DSP attacks across datasets.
Attacks: GAU = Gaussian noise (SNR=60), AMP = Amplitude scaling (0.5), LPF = Low-pass filtering (4kHz), and RSM = Resampling (16kHz).}
\label{tab:dsp_attack}
\centering
\scriptsize
\setlength{\tabcolsep}{3.2pt}
\resizebox{\columnwidth}{!}{%
\begin{tabular}{ll|cccc}
\toprule
Dataset & Attack & AudioSeal~\cite{audioseal_2024} & SilentCipher~\cite{singh2024silentcipher} & WavMark~\cite{wavmark_2024} & \textbf{Latent-Mark} \\
\midrule
\multirow{4}{*}{\textbf{AIR}} & GAU & \textbf{100.00} & ~~\underline{18.60} & ~~~~3.33 & \textbf{100.00} \\
 & AMP & \textbf{100.00} & ~~~~0.00 & ~~~~\underline{3.49} & \textbf{100.00} \\
 & LPF & \textbf{100.00} & ~~\underline{10.17} & ~~~~3.49 & \textbf{100.00} \\
 & RSM & \textbf{100.00} & ~~~~\underline{9.30} & ~~~~3.49 & \textbf{100.00} \\
\cline{1-6}
\multirow{4}{*}{\textbf{Freischuetz}} & GAU & ~~68.49 & ~~\underline{97.26} & \textbf{100.00} & ~~69.86 \\
 & AMP & \textbf{100.00} & 0.00 & \textbf{100.00} & ~~\underline{71.23} \\
 & LPF & \textbf{100.00} & ~~\underline{93.15} & \textbf{100.00} & ~~56.16 \\
 & RSM & \textbf{100.00} & ~~\underline{95.89} & \textbf{100.00} & ~~64.38 \\
\cline{1-6}
\multirow{4}{*}{\textbf{GuitarSet}} & GAU & ~~\underline{90.00} & \textbf{100.00} & \textbf{100.00} & ~~67.78 \\
 & AMP & \textbf{100.00} & ~~~~0.00 & \textbf{100.00} & ~~\underline{91.11} \\
 & LPF & \textbf{100.00} & ~~\underline{94.44} & \textbf{100.00} & \textbf{100.00} \\
 & RSM & \textbf{100.00} & ~~\underline{92.22} & \textbf{100.00} & \textbf{100.00} \\
\cline{1-6}
\multirow{4}{*}{\textbf{jaCappella}} & GAU & ~~20.00 & ~~46.00 & ~~\underline{84.00} & \textbf{100.00} \\
 & AMP & \textbf{100.00} & ~~~~0.00 & ~~84.00 & ~~\underline{88.00} \\
 & LPF & \textbf{100.00} & ~~38.00 & ~~\underline{84.00} & ~~78.00 \\
 & RSM & \textbf{100.00} & ~~34.00 & ~~\underline{84.00} & ~~\underline{84.00} \\
\cline{1-6}
\multirow{4}{*}{\textbf{LibriSpeech}} & GAU & \textbf{100.00} & ~~\underline{99.19} & \textbf{100.00} & \textbf{100.00} \\
 & AMP & \textbf{100.00} & ~~\underline{98.39} & \textbf{100.00} & \textbf{100.00} \\
 & LPF & \textbf{100.00} & ~~\underline{89.52} & \textbf{100.00} & ~~75.81 \\
 & RSM & \textbf{100.00} & ~~\underline{91.13} & \textbf{100.00} & \textbf{100.00} \\
\bottomrule
\end{tabular}%
}
\end{table}

\subsection{Robustness to Prior Attacks}
\label{sec:dsp_attack}
\lily{Besides imperceptibility, we extend our comparisons to evaluate the robustness of Latent-Mark against traditional digital signal processing (DSP) distortions, benchmarking against the same three baselines: AudioSeal, WavMark, and SilentCipher. Following the evaluation protocol outlined in SoK~\cite{wen2025sok}, we subject the watermarked audio to four highly diverse signal distortion attacks to cover a wide spectrum of degradation: Gaussian noise, amplitude scaling, low-pass filtering, and resampling. The hyperparameters are set consistently with prior work.}

\lily{As shown in Table~\ref{tab:dsp_attack}, AudioSeal consistently achieves the highest robustness across these traditional distortions, which is expected given that it is explicitly trained on augmented audio editing data to ensure resilience against such modifications. Meanwhile, Latent-Mark and WavMark exhibit competitive, dataset-dependent performance; our method outperforms WavMark on the AIR and jaCappella datasets, whereas WavMark holds an advantage on Freischuetz and GuitarSet. This indicates that while Latent-Mark is primarily designed to survive neural codec bottlenecks, it successfully retains robustness against conventional attacks, performing on par with dedicated robust watermarking methods. Finally, SilentCipher yields the lowest detection rates in this setting---dropping to as low as 0.00\% under amplitude scaling attacks across several datasets---which we attribute to its reliance on delicate temporal and phase alignments that are easily disrupted by broad signal distortions.}


\vspace{5pt}
\noindent
\lily{\textbf{Summary of Trade-offs.} Concluding from Sections~\ref{sec:imperceptibility} and ~\ref{sec:dsp_attack}, Latent-Mark establishes a balance between acoustic transparency and robustness. While AudioSeal and WavMark exhibit strong resilience to traditional DSP distortions, they do so at the severe expense of audio quality and completely fail under neural codec compression. Conversely, while SilentCipher maintains high imperceptibility, it remains vulnerable to both DSP attacks and codec bottlenecks. Latent-Mark delivers acoustic fidelity comparable to the highly constrained SilentCipher, while uniquely surviving the extreme bottleneck of neural codec compression and maintaining competitive defense against standard DSP perturbations.}

\section{Conclusion}
\label{sec:conclusion}

We propose Latent-Mark, the first zero-bit audio watermarking framework specifically engineered to survive neural codec compression, a process that causes catastrophic failure in traditional methods. Using the insight that watermark information must be embedded directly into the 
latent space \update{manifolds} to 
\update{survive}
quantization bottlenecks, we successfully bridge the gap between DSP robustness and neural codec survivability. 

Our findings indicate that latent-space alignment is critical in robustness towards neural codec compression; specifically, perturbations guided by the codebook's topological clusters achieve higher imperceptibility while maintaining high detection accuracy. Beyond its specialized resilience to neural synthesis, \textit{Latent-Mark} achieves competitive performance in both perceptual transparency and robustness against DSP attacks. We show that black-box transferability can be achieved by jointly optimizing across diverse codec architectures, and we hope our work will inspire future research on unified watermarking methods to adapt to the evolving landscape of generative neural synthesis.

\section{Acknowledgment}
This work was supported in part by the National Science and Technology Council under Grants NSTC 114-2634-F-002-004 and NSTC 114-2634-F-002-003-MBK.


\section{References}

\makeatletter
\renewcommand{\bibsection}{}
\makeatother

\bibliographystyle{IEEEtran}
\bibliography{mybib}

\end{document}